\documentclass[a4paper,fleqn]{cas-dc}

\usepackage[authoryear,longnamesfirst]{natbib}
\usepackage{lineno,hyperref}
\usepackage{ulem}
\usepackage{algorithm2e,setspace}
\usepackage{subfigure}
\usepackage{multirow}
\usepackage{bbding}
\usepackage{mathrsfs}
\usepackage{amsthm,amsmath,amssymb}
\usepackage{tabularx}
\usepackage{natbib}
\usepackage{threeparttable}

\def\tsc#1{\csdef{#1}{\textsc{\lowercase{#1}}\xspace}}
\tsc{WGM}
\tsc{QE}
\tsc{EP}
\tsc{PMS}
\tsc{BEC}
\tsc{DE}

\begin{document}
\let\WriteBookmarks\relax
\def\floatpagepagefraction{1}
\def\textpagefraction{.001}

\shorttitle{}

\shortauthors{Xie et~al.}

\title [mode = title]{Advancing Robust Underwater Acoustic Target Recognition through Multi-task Learning and Multi-Gate Mixture-of-Experts}                      




%
\author[address1,address2]{Yuan Xie}[
                        style=chinese,
                        orcid=0000-0003-3803-0929]


\ead{xieyuan@hccl.ioa.cn.cn}



\credit{Conceptualization, Methodology, Software, Validation, Formal analysis, Investigation, Data Curation, Writing - Original Draft, Writing - Review \& Editing, Visualization}

\author[address1,address2]{Jiawei Ren}[style=chinese]
\ead{renjiawei@hccl.ioa.cn.cn}
\credit{Methodology, Investigation}

\author[address1,address2]{Junfeng Li}[style=chinese]
\ead{lijunfeng@hccl.ioa.cn.cn}
\credit{Data Curation, Supervision}

\author[address1,address2,address3]{Ji Xu}[style=chinese, orcid=0000-0002-3754-228X]
\ead{xuji@hccl.ioa.cn.cn}
\cormark[1]

\credit{Resources, Writing - Review \& Editing, Supervision, Project administration, Funding acquisition}

\affiliation[address1]{organization={Key Laboratory of Speech Acoustics and Content Understanding, Institute of Acoustics, Chinese Academy of Sciences},
    addressline={No.21, Beisihuan West Road, Haidian District},
    postcode={100190},
    city={Beijing},
    country={China}}
    
\affiliation[address2]{organization={University of Chinese Academy of Sciences},
    addressline={No.80, Zhongguancun East Road, Haidian District}, 
    postcode={100190},
    city={Beijing},
    country={China}}

\affiliation[address3]{organization={State Key Laboratory of Acoustics, Institute of Acoustics, Chinese Academy of Sciences},
    addressline={No.21, Beisihuan West Road, Haidian District},
    postcode={100190},
    city={Beijing},
    country={China}}

\cortext[cor1]{Corresponding author}



\begin{abstract}
Underwater acoustic target recognition has emerged as a prominent research area within the field of underwater acoustics. However, the current availability of authentic underwater acoustic signal recordings remains limited, which hinders data-driven acoustic recognition models from learning robust patterns of targets from a limited set of intricate underwater signals, thereby compromising their stability in practical applications. To overcome these limitations, this study proposes a recognition framework called M3 (Multi-task, Multi-gate, Multi-expert) to enhance the model's ability to capture robust patterns by making it aware of the inherent properties of targets. In this framework, an auxiliary task that focuses on target properties, such as estimating target size, is designed. The auxiliary task then shares parameters with the recognition task to realize multi-task learning. This paradigm allows the model to concentrate on shared information across tasks and identify robust patterns of targets in a regularized manner, thereby enhancing the model's generalization ability. Moreover, M3 incorporates multi-expert and multi-gate mechanisms, allowing for the allocation of distinct parameter spaces to various underwater signals. This enables the model to process intricate signal patterns in a fine-grained and differentiated manner. To evaluate the effectiveness of M3, extensive experiments were implemented on the ShipsEar underwater ship-radiated noise dataset. The results substantiate that M3 has the ability to outperform the most advanced single-task recognition models, thereby achieving the state-of-the-art performance.
\end{abstract}





\maketitle


\section{Introduction}
Underwater acoustic target recognition is a crucial component of marine acoustics~\cite{rajagopal1990target}. Its purpose is to automatically recognize different types of underwater targets by analyzing their radiating sound. This technology finds extensive applications in underwater surveillance~\cite{sutin2010stevens}, marine resources development and protection~\cite{vaccaro1998past}, and security defense~\cite{fillinger2010towards}.

In recent years, deep learning has emerged as the dominant technique for underwater acoustic recognition systems~\cite{niu2023advances,hummel2024survey}. Data-driven acoustic recognition models based on deep learning have shown promising performance on publicly available datasets~\cite{santos2016shipsear, irfan2021deepship}. These data-driven recognition models may struggle to capture underlying robust patterns from limited but complicated signal recordings, leading to issues such as overfitting and limited generalization capabilities in real-world ocean scenarios~\cite{xie2023advancing, irfan2021deepship, xie2022underwater}. Furthermore, the difficulties and costs associated with underwater signal acquisition, as well as restrictions imposed by security defense and military applications ~\cite{santos2016shipsear, xie2024unraveling}, may perpetuate the scarcity of publicly accessible underwater acoustic signal recordings, hindering the development of large-scale robust models over an extended period~\cite{xu2023underwater}. To make further progress under such conditions, a promising idea is to enhance the model's capacity to capture robust patterns by making it aware of targets' inherent properties (e.g., target size).

According to our research, the majority of existing literature in underwater acoustic recognition~\cite{xie2022adaptive, erkmen2008improving, simonovic2021acoustic, jia2022deep} relies solely on category labels as supervisory information, which provides limited insights into the intrinsic patterns of targets when data is insufficient. In response to this aspect, recent advancements in machine learning techniques have offered opportunities to exploit additional information beyond category labels. Among them, multi-task learning~\cite{caruana1997multitask, ma2018modeling, misra2016cross, tang2020progressive} can serve as an attractive solution. For acoustic signal recognition, the multi-task model can consider exploiting intrinsic properties from signals as the auxiliary task, which shares parameters with the recognition task. The interaction between tasks facilitates mutual information transfer and initiates regularization constraints. This can prompt the model to focus on common robust features of target signals across different perspectives, while disregarding task-irrelevant noise. Such a paradigm exhibits enhanced data utilization efficiency and can assist the model in acquiring greater robustness.

In this study, we employ a multi-task learning framework to enable the recognition task to perceive robust patterns related to the inherent properties of targets in underwater signals, with ``target size estimation'' serving as the auxiliary task. This auxiliary task does not require additional annotations since the target size can be easily inferred based on the category labels. In addition, it can also encourage the model to learn about size-related acoustic characteristics such as resonant frequencies, hull noise~\cite{song2014reduction}, etc., thus improving the model's insights into signals. Moreover, inspired by recent work on the Mixture of Experts (MoE)~\cite{jacobs1991adaptive,riquelme2021scaling,xie2024unraveling} and the multi-gate mechanism~\cite{ma2018modeling,tang2020progressive}, we adopt an improved multi-task framework called M3 (\textbf{M}ulti-task, \textbf{M}ulti-gate, \textbf{M}ulti-expert) to fully exploit the potential of multi-task learning. Specifically, M3 employs multiple independent network layers, referred to as expert layers, to replace the traditional shared layer to facilitate information sharing between tasks. Expert layers have identical architectures but distinct parameters, allowing them to specialize in different aspects and provide fine-grained knowledge with separate parameter spaces. Additionally, multiple gating layers~\cite{jacobs1991adaptive,ma2018modeling} are employed to dynamically learn task-specific weights, allowing each task to linearly add the outputs of expert layers with unique weights and acquire task-specific representations (refer to Sections III.B and III.C for detailed introductions to the expert layer and the gating layer). M3 also incorporates one-dimensional frequency domain features as the input to the gating layers, replacing conventional inputs, to provide low-dimensional and non-redundant information. Furthermore, this study presents an optional strategy to alleviate interference among tasks by designating certain experts exclusively for specific tasks. To validate the superior performance of the multi-task framework and M3, a series of experiments were implemented on the ShipsEar dataset~\cite{santos2016shipsear}. The results demonstrate that the multi-task models can achieve superior recognition accuracy, with M3 exhibiting a notable performance enhancement. On the ShipsEar dataset, M3 can achieve a remarkable accuracy of 87.07$\pm$2.43\% on the 9-class recognition task, reaching state-of-the-art performance. The main contributions of this study can be summarized as follows:


\begin{itemize}

\item Identifying the limitations of current acoustic recognition methods under existing data conditions and employing multi-task learning to enhance the model's ability to capture robust patterns by leveraging inherent properties of targets.

\item Introducing multi-expert and multi-gate mechanisms, which facilitate the processing of complex underwater acoustic signals by utilizing specialized and separate parameter spaces.

\item Proposing two optimization strategies: incorporating low dimensional features as input to the gating layer, and converting certain expert layers into task-specific ones.

\item Conducting abundant experiments to validate the superiority of the M3 model, which achieves state-of-the-art performance on the ShipsEar dataset.

\end{itemize}

\section{Related Works}

\subsection{Deep Learning-Based Underwater Acoustic Target Recognition}

With the advent of deep learning and the accumulation of underwater noise databases~\cite{santos2016shipsear,irfan2021deepship}, related research based on deep neural networks has become increasingly prevalent in underwater acoustic target recognition~\cite{li2022combined,xie2023guiding,feng2024underwater}. Especially in recent years, the number of relevant studies has experienced a rapid increase. As reported in the literature, many studies focused on optimizing the design of acoustic features, such as multi-scale spectral~\cite{jiang2020multi}, CMS-based modulation spectrum~\cite{luo2021underwater}, learnable Gabor filter banks~\cite{ren2022ualf}, cepstrum-wavelet~\cite{jia2022deep}, adaptive wavelet spectrum~\cite{xie2022adaptive}, improved EMD~\cite{jin2023offshore}, 3D dynamic MFCC feature~\cite{yang2023lightweight}, 3D spectrogram~\cite{tang2023differential}, etc. These well-designed acoustic features, which contain temporal or spectral information with distinct focal points, have demonstrated their effectiveness when utilized as input features for deep neural networks. In addition, there is also a lot of work devoted to optimizing the neural network architecture~\cite{tang2023differential,yang2023lightweight,xie2024unraveling}, or using neural networks to implement adaptive representation extraction~\cite{ji2023underwater,kamalipour2023passive}, data augmentation~\cite{li2023data,xu2023underwater}, noise reduction~\cite{zhou2020denoising,zhou2023novel}, and other techniques to achieve better recognition performance.

Whereas, the majority of existing studies in the field of underwater acoustic target recognition solely rely on category labels as supervisory information. It provides limited insights into the intrinsic patterns of targets, particularly when data is insufficient~\cite{xie2022underwater}. Several studies have acknowledged the importance of considering factors beyond category labels, such as acoustic channels~\cite{li2023data}, source distance, channel depth, and wind speed~\cite{xie2022underwater}. However, the training paradigm utilized in those studies, which involved techniques such as data augmentation and contrastive learning, lacks high data utilization efficiency, leaving room for further optimization.

\subsection{Multi-Task Learning}
Multi-task learning (MTL) is a methodology that enables learning across multiple tasks by simultaneously solving multiple objectives. By leveraging parameter sharing, MTL can learn commonalities and differences across tasks, thus enhancing data utilization efficiency and generalization capabilities for each individual task~\cite{ruder2017overview}. The most commonly used and fundamental MTL structure is based on hard parameter sharing~\cite{caruana1997multitask}. On this basis, some improved MTL algorithms have also been proposed, including the cross-stitch network~\cite{misra2016cross}, sluice network~\cite{ruder2017sluice}, etc.


In recent years, several studies have explored the use of mixture of experts (MoE)~\cite{jacobs1991adaptive, riquelme2021scaling, xie2023moec} for implementing MTL with dynamic weights. Compared with fundamental MTL structures, the MoE-based MTL has a dynamic parameter-sharing mechanism, which reduces the contradiction between tasks and shared parameters. The representative work that applies MoE to MTL includes the multi-gate MoE~\cite{ma2018modeling}, progressive layered extraction (PLE)~\cite{tang2020progressive}, differentiable select-k (DSelect-k)~\cite{hazimeh2021dselect}, etc.

\subsection{Multi-Task Learning in Underwater Acoustics}

In the field of underwater acoustics, MTL has a wide range of applications. For instance, underwater acoustic communication-related work tends to utilize extra tasks, including channel estimation~\cite{liang2023multitask}, channel tracking~\cite{zhang2022complex}, channel equalization~\cite{stojanovic1993adaptive}, demodulation~\cite{zhang2021meta}, etc., to improve performance. Besides, other areas of underwater acoustics, such as synthetic aperture sonar (SAS) image classification~\cite{williams2019transfer,gerg2020data}, sound source localization~\cite{liu2020multi,wu2021sound}, also often employ MTL to enhance performance. However, for underwater acoustic target recognition, the investigation of MTL is still in its nascent stage. Only a limited amount of research has been dedicated to this area. According to our survey, Zeng et, al.~\cite{zeng2020multi} proposed a multi-task sparse feature learning method for underwater acoustic target recognition by recovering and enhancing prominent structures on spectra; Li et, al.~\cite{li2023robust} designed an anti-noise task and frequency-selection task to optimize the acoustic feature extraction. We note that most previous works in this field used MTL to optimize the extraction or learning of acoustic features, and did not enable the model to perceive knowledge related to the robust properties of targets directly. In addition, these methods only used the fundamental MTL framework, which has considerable room for improvement. In this study, we adopt ``target size estimation'' as the auxiliary task in pursuit of the perception of the targets' inherent properties, and implement a MoE-style MTL approach in the domain of underwater acoustic recognition.


\section{Methods}

\subsection{Preprocessing and Feature Extraction}

First, the underwater acoustic signals are collected by underwater hydrophones or hydrophone arrays, and then stored in \emph{wav} format files. After that, the single-channel audio files are fed into a bandpass filter with a wide bandwidth of 10~Hz -- 26360~Hz. The frequency range is determined based on the Nyquist sampling theorem and the cutoff frequency of the recording devices~\cite{santos2016shipsear}. Additionally, we apply the mean-variance normalization (z-score normalization) and pre-emphasis techniques to filtered waveforms to enhance the signal components and suppress interference.

\begin{figure*}
    \centering
    \includegraphics[width=1\linewidth]{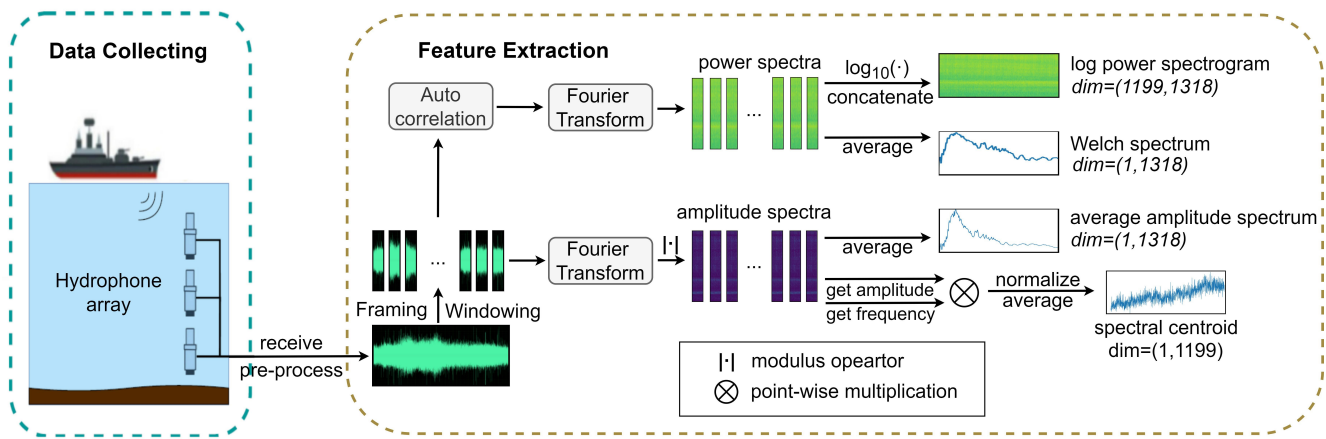}
    \caption{The overall process of data collecting and feature extraction. The extracted acoustic features consist of the 2-D main feature -- log power spectrogram, along with three candidate 1-D gating features. The dimensions of the extracted features are also presented in the figure.}
    \label{fig_fea}
    \vspace{-2px}
\end{figure*}

After preprocessing, feature extraction is performed to convert signals into pre-designed acoustic features. The overview of the feature extraction process employed in this study is presented in Fig.~\ref{fig_fea}. Two types of features are extracted throughout the process. The first type, referred to as the ``main feature'', serves as the input for the main body of the model. The other feature, known as the ``gating feature'', is used as the input only for the gating layers. The different usages of main features and gating features require them to have different characteristics and emphases.

The main feature is used to provide abundant time-frequency information for model training. In this study, the 2-D log power spectrogram, which has shown promising performance on underwater acoustic recognition tasks~\cite{luo2021underwater,luo2021underwaterv2}, is extracted as the main feature. The extraction process is illustrated in Fig.~\ref{fig_fea}. The signal is first framed with a frame length of 50~ms and a frame shift of 25~ms. Subsequently, the Hanning window is applied to each short-time frame to mitigate spectrum leakage. Afterward, the auto-correlation and Fourier transform are performed on the windowed frames to obtain the power spectra. The power spectra of all frames are concatenated along the temporal dimension, and a logarithm operation is then applied to derive the log power spectrogram. The log power spectrogram contains rich time-frequency information, and its logarithmic value highlights the intensity discrepancy of different frequency components and enhance the distinguishability of features. In Section V.A, preliminary experiments are presented to demonstrate the superiority of the log power spectrogram over other optional features (e.g., DEMON spectrum, Mel spectrogram, CQT spectrogram, amplitude spectrogram, and power spectrogram) in the underwater acoustic recognition task.

The gating features serve as the inputs for the gating layer, which is responsible for providing task-specific weights for expert layer outputs. In MoE-style models~\cite{ma2018modeling, riquelme2021scaling, xie2023moec}, the architecture of the gating layer should be simple to avoid intricate and non-intuitive weight assignment. Therefore, gating features need to be non-redundant and low-dimensional to avoid underfitting on the simple gating layer. Since the gating input of traditional MoE-style models seems to be inapposite in underwater acoustic recognition (detailed discussion is provided in Section III.B), in this study, three 1-D frequency-domain features -- Welch spectrum~\cite{kang2004underwater}, average amplitude spectrum, and spectral centroid~\cite{li2021underwater} are extracted as candidate gating features. The relevant extraction steps are also shown in Fig.~\ref{fig_fea}, where the frame lengths, frame shifts, and window functions used for extracting these gating features follow the aforementioned settings. The extraction steps for these three candidate gating features are specified as follows: First, the Welch spectrum is obtained by averaging the power spectra of all frames, as outlined in the preceding paragraph. Secondly, to extract the average amplitude spectrum, the short-time Fourier transform (STFT) and modulus operator are employed to obtain the amplitude spectrum of each frame. Then, the spectra from all frames are averaged to derive the average amplitude spectrum. The extraction of the spectral centroid is also conducted on amplitude spectra. The spectral centroid for each frame is obtained by summing the products of spectra components (amplitudes) and their respective frequencies, followed by normalizing this summation by dividing it by the total sum of amplitudes. Finally, the spectral centroids for all frames are concatenated along the temporal dimension to obtain the spectral centroid feature.

\subsection{Further Analysis on Gating Features}



The gating feature of traditional MoE-style models is consistent with the input of expert layers~\cite{ma2018modeling,riquelme2021scaling}. However, such a design is unsuitable for underwater acoustic tasks, where the input of expert layers is high-dimensional spectrograms. This not only brings a lot of additional computational consumption, but also easily causes the gating layer to fall into under-fitting~\cite{xie2023moec}, thus affecting the rationality of the gating layer output. To address these limitations, we adopt 1-D frequency-domain features as gating features. These features contain non-redundant and task-related patterns (e.g., line spectral frequencies), which are more suitable to serve as the gating feature in underwater acoustic tasks. Moreover, these features offer the advantage of being low-dimensional, which enhances computational efficiency.

\begin{figure*}
    \centering
    \includegraphics[width=0.9\linewidth]{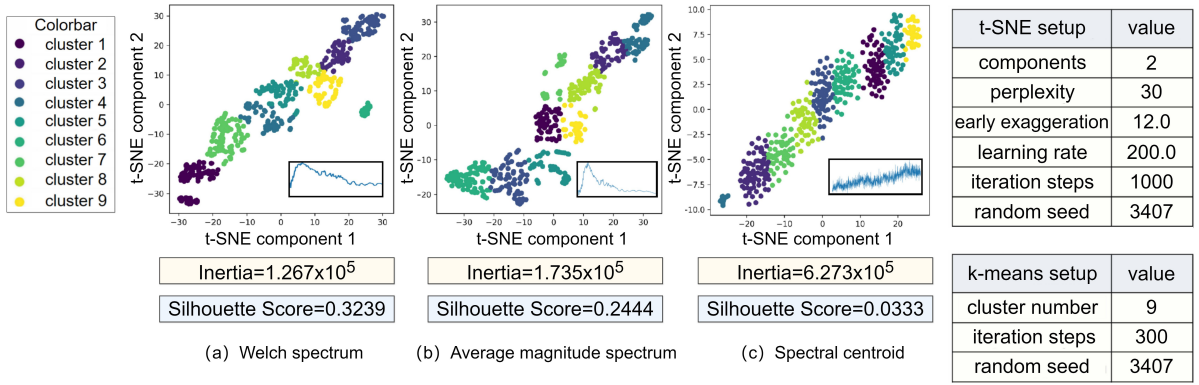}
    \caption{The distribution of three candidate gating features: (a) Welch spectrum, (b) average amplitude spectrum, (c) spectral centroid, on the ShipsEar dataset. The visualization results were derived using t-SNE and k-means clustering. This figure also provides the parameter setups for t-SNE and k-means algorithms, along with the associated values of clustering inertia and Silhouette score.}
    \label{fig_gate_input}
    \vspace{-2px}
\end{figure*}

To select the optimal gating feature, the t-SNE (t-distributed stochastic neighbor embedding) algorithm~\cite{van2008visualizing} and k-means clustering algorithm~\cite{macqueen1967some} were employed to visualize the distribution of three candidate gating features on ShipsEar dataset. More specifically, gating features were initially reduced to a two-dimensional space by the t-SNE algorithm to facilitate visualization, and then these data points were clustered into a specified number of clusters using the k-means algorithm to reveal the underlying data structure (relevant parameter setups for both algorithms are illustrated in the right portion of Fig.~\ref{fig_gate_input}). We aspire for the gating features to possess distinguishable distributions, thereby facilitating the gating layer to effectively capture signal patterns. As can be seen from Fig.~\ref{fig_gate_input}, the distribution of the average amplitude spectrum (Fig.~\ref{fig_gate_input} (b)) exhibits a more dispersed pattern, while the distribution for the spectral centroid (Fig.~\ref{fig_gate_input} (c)) displays large intra-cluster distances while maintaining small inter-cluster distances. In contrast, the distribution for the Welch spectrum (Fig.~\ref{fig_gate_input} (a)) exhibits small intra-class distances and appropriate inter-class distances, which showcases the best distinguishability among the candidate gating features. In addition, quantitative metrics on clustering quality -- inertia (cluster sum of square) and Silhouette Score~\cite{shahapure2020cluster} are also reported. In general, smaller inertia and larger Silhouette Score represent better clustering effects, which indicate a better distribution status. As can be seen from Fig.~\ref{fig_gate_input}, the two clustering metrics for Welch spectra are obviously superior to those of the other two features. Based on the visual analyses and quantifiable results, along with the subsequent experimental results presented in Section V.A, we consider the 1-D Welch spectrum as the optimal gating feature.

\subsection{Framework and Architecture of M3 and M3-TSE}


\begin{figure*}
    \centering
    \includegraphics[width=1.0\linewidth]{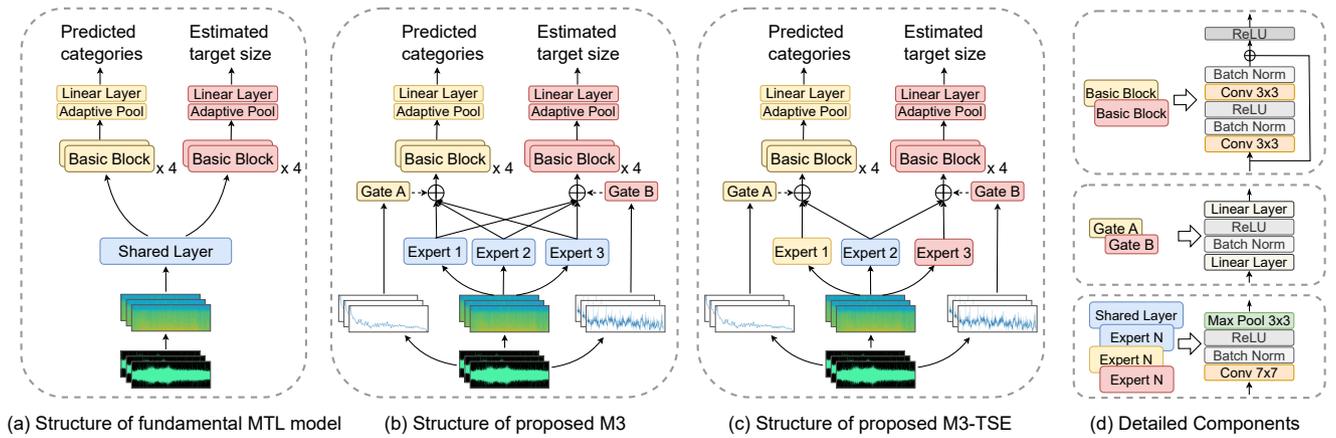}
    \caption{An overview of the structure of (a) fundamental MTL model, (b) M3, (c) M3-TSE, (d) detailed architecture components. In this figure, yellow represents the relevant components of the recognition task, red represents the relevant components of the target size estimation task, and blue represents the shared parts between the two tasks.}
    \label{fig2}
    \vspace{-2px}
\end{figure*}

This subsection presents the framework and architecture of our proposed M3 and its optimized version, M3-TSE (M3 with task-specific experts). The framework of M3 and M3-TSE is depicted in Fig.~\ref{fig2} (b) and (c). Both models share a similar network framework, which includes multiple expert layers, two gating layers, and two tower networks. While M3 shares all experts for both tasks, M3-TSE retains specific experts for each task while sharing others. The two tasks share parameters and knowledge through the shared expert layers.

\begin{table*}[ht]
\normalsize
    \centering
    \caption{The detailed network architectures for M3 and M3-TSE, where ``expert num'' and ``kind num'' are task-related parameters. The term ``Conv2d'' denotes the two-dimensional convolutional layer.}
	\scalebox{0.75}{\begin{tabular}{ll}
		\hline
		  Module&Detailed network architecture\\ 
            \hline
            Expert layer \& Shared layer&Conv2d(in channel=1, out channel=64, kernel size=7, stride=2, padding=3)\\ 
             &Batch Normalization 2d(num features=64)\\ 
             &ReLU()\\ 
             &Max pooling(kernel size=3, stride=2, padding=1)\\
             \hline
            Gating layer&Linear(in features=1318, out features=128)\\
            &Batch Normalization 1d(num features=128)\\
            &ReLU()\\
            &Linear(in features=128, out features=expert num)\\
            \hline
            Tower network&Basic block(64,64), Basic block(64,64)\\
             &Basic block(64,128),Basic block(128,128)\\
             &Basic block(128,256), Basic block(256,256)\\
             &Basic block(256,512), Basic block(512,512)\\
             &Adaptive average pooling(output size=(1, 1))\\
             &Linear(in features=512, out features=kind num)\\
            \hline
            Basic block(in dim, out dim) & Conv2d(in dim, out dim, kernel size=3, padding=1)\\
            &Batch Normalization 2d(num features=out dim)\\
            &ReLU()\\
            &Conv2d(out dim, out dim, kernel size=3, padding=1)\\
            &Batch Normalization 2d(num features=out dim)\\
            
		\hline
        \label{tab1}
	\end{tabular}}
\end{table*}

The detailed architecture of all network layers can be found in Table~\ref{tab1}. M3 and M3-TSE both contain multiple expert layers, while shared and specific experts have the same architecture, comprising a 7$\times$7 convolutional layer, a batch normalization (BN) layer, a ReLU layer, and a 3$\times$3 max pooling layer (the architecture of the expert layer is also shown in Fig.~\ref{fig2} (d)). Each expert possesses different convolutional layer parameters, enabling it to perceive and acquire distinct knowledge aspects from various input features. Moreover, each model contains two task-specific gating layers, responsible for adaptively learning weights for the main task (i.e. target category recognition) and the auxiliary task (i.e. target size estimation), respectively. The two gating layers have independent parameters but share the same architecture. Each gating layer consists of a simple multi-layer perceptron (MLP), which contains two linear layers, a BN layer, and a ReLU layer. The output dimensions of the last linear layer can be adjusted based on the number of experts.

Similar to the gating layers, the tower network is also task-specific. They are responsible for transforming the weighted representations into predicted results for their respective task. Each tower network follows the architecture of ResNet~\cite{he2016deep} and adopts the four residual layers of ResNet-18. Each residual layer contains a stack of two basic blocks, comprising two 3$\times$3 convolutional layers, two BN layers, a ReLU layer, and a skip connection (the architecture of the basic block is also shown in Fig.~\ref{fig2}(d)). Additionally, each tower network includes an adaptive average pooling layer and a task-related linear layer. The output dimension of the linear layer is determined by the number of categories to be predicted for the given task.

\subsection{Detailed Workflow of M3 and M3-TSE}

Then, this subsection introduces the detailed workflow of our proposed M3 and ME-TSE. Denote the 2-D main feature as $x$, and use the $\mathrm{main}$ and $\mathrm{aux}$ subscripts to represent the specific components for the main and auxiliary tasks. For example, there are task-specific labels $y_{\mathrm{main}}$ (target type), $y_{\mathrm{aux}}$ (target size), gating features $s_\mathrm{main}, s_\mathrm{aux}$, gating layer $G_\mathrm{main}(\cdot), G_\mathrm{aux}(\cdot)$, and tower network $T_\mathrm{main}(\cdot), T_\mathrm{aux}(\cdot)$. The workflow of M3 is briefly shown in Fig.~\ref{fig2}(b). First, $x$ is fed into $N$ expert layers $E_1(\cdot),..., E_N(\cdot)$, and get corresponding outputs $r_i=E_i(x)\in \mathbb{R}^{B \times 64\times 300\times 330} (i=1,2,...,N)$, where B represents the batch size, 64 represents the number of channels, and 300$\times$330 represent the dimensions of the feature map after convolution and pooling (the dimension of the feature map will change as the dimension of input features change). These outputs can be stacked into a matrix $r=[r_1,r_2,...,r_N]\in \mathbb{R}^{B \times N\times64\times 300\times 330}$, which contains the outputs of all experts. Afterwards, for the gating module, the gating features $s_\mathrm{main}, s_\mathrm{aux}$ are fed into their corresponding gating layer $G_\mathrm{main}(\cdot), G_\mathrm{aux}(\cdot)$ to obtain task-specific weight matrix $w_\mathrm{main}, w_\mathrm{aux} \in \mathbb{R}^{B \times N}$. 

\begin{equation}
\begin{aligned}
    &w_\mathrm{main} = \mathrm{softmax}(G_\mathrm{main}(s_\mathrm{main})),\\
    &w_\mathrm{aux} = \mathrm{softmax}(G_\mathrm{aux}(s_\mathrm{aux})),
\end{aligned}
\end{equation}

where $\mathrm{softmax}(p_i) = e^{p_i} / \sum_{j=1}^{N} e^{p_j}$ is the function that transforms input vectors into categorical probabilities. It should be mentioned that the softmax function here is applied to the last dimension of the input vector. In the weight matrix $w$, the $w_{i,*}$ ($i,*$ denotes the $i$-th row of the matrix) represents the weights for each expert. Then, the weight matrix $w$ and the representation matrix $r$ are point-wise multiplied, and then linearly added to obtain the weighted representation $r_\mathrm{main}, r_\mathrm{aux}$:

\begin{equation}
\begin{aligned}
    r_\mathrm{main}
    &=\mathrm{sum}(r\cdot w_\mathrm{main})
    &=\sum_{i=1}^{N} r_i\cdot w_{\mathrm{main}\,{i,*}},\\
    r_\mathrm{aux}
    &=\mathrm{sum}(r\cdot w_\mathrm{aux})
    &=\sum_{i=1}^{N} r_i\cdot w_{\mathrm{aux}\,{i,*}}.\\
\end{aligned}
\end{equation}

Then, the weighted representations $r_\mathrm{main},r_\mathrm{aux}$ are fed into task-specific back-end tower networks and get corresponding outputs: $z_\mathrm{main}=T_\mathrm{main}(r_\mathrm{main}) \in \mathbb{R}^{B \times C_\mathrm{main}}$, $z_\mathrm{aux}=T_\mathrm{aux}(r_\mathrm{aux})\in \mathbb{R}^{B \times C_\mathrm{aux}}$, where $C_\mathrm{main}$ and $C_\mathrm{aux}$ represent the number of categories to be predicted for the main task and auxiliary task. Then, each task separately calculates the cross entropy loss between the output of the tower network $z_\mathrm{main},z_\mathrm{aux}$ and true labels $y_{\mathrm{main}},y_{\mathrm{aux}}$. The final joint loss $\mathcal{L}$ can be obtained by linearly summing the cross-entropy losses of both tasks, and the optimization goal is to minimize the final joint loss. The formula of the loss function is presented as follows:

\begin{small}
\begin{equation}
\begin{aligned}
    \mathcal{L}
    &=\mathcal{L}_\mathrm{main} + \alpha \cdot \mathcal{L}_\mathrm{aux}\\
    &=-\frac{1}{B} (\sum_{i=1}^{B}\sum_{j=1}^{C_\mathrm{main}}y_{\mathrm{main}\,{i,j}}\cdot\log(z_{\mathrm{main}\,{i,j}})
    +\alpha \sum_{i=1}^{B}\sum_{j=1}^{C_\mathrm{aux}}y_{\mathrm{aux}\,{i,j}}\cdot\log(z_{\mathrm{aux}\,{i,j}})),
\end{aligned}
\end{equation}
\end{small}

where $\alpha$ is an adjustable coefficient used to control the weight between tasks. The model requires a moderate weight to balance the two tasks, so the range of $\alpha$ values in this study is limited to $[0.1,5]$. In addition, the homoscedastic uncertainty algorithm~\cite{kendall2018multi,neilsen2021learning} is also employed in pursuit of an optimal $\alpha$ value obtained through adaptive learning (see Section V.C for related experiments on $\alpha$). Then, the gradient of the final joint loss $\mathcal{L}$ can be calculated and all the learnable parameters in the model can be updated by the backpropagation algorithm.

The above is the workflow of M3. Building upon this, an improved version called M3-TSE, as shown in Fig.~\ref{fig2}(c), is proposed to alleviate the distractions between tasks. The expert layer of M3-TSE consists of shared experts and task-specific experts. Each task can only access shared experts and their respective specific experts. Let $N_\mathrm{sp}$ denote the number of specific experts for a given task, and $N_\mathrm{sh}$ represent the number of shared experts. The outputs of $N_\mathrm{sp}$ specific experts and $N_\mathrm{sh}$ shared experts are stacked into a matrix, denoted as $r\in \mathbb{R}^{B \times (N_\mathrm{sh}+N_\mathrm{sp})\times64\times 300\times 330}$. Subsequently, the gating layer adjusts the output dimension to $N_\mathrm{sh}+N_\mathrm{sp}$ and generates the weight matrix $w\in \mathbb{R}^{B \times (N_\mathrm{sh}+N_\mathrm{sp})}$ according to Equation (1). Then the task-specific representations can be calculated using $r$ and $w$ according to Equation (2). After that, the subsequent procedures are exactly the same as those for M3. By incorporating the task-specific mechanism into the expert layers, M3-TSE effectively mitigates the distractions between tasks~\cite{tang2020progressive} by further promoting uniqueness across tasks.

\section{Experiment Setups}


\subsection{Datasets}
In this study, all experiments were implemented on an underwater ship-radiated noise dataset -- ShipsEar~\cite{santos2016shipsear}. ShipsEar is an open-source database of underwater recordings of ship-radiated sounds. The recordings were collected in different areas of the Atlantic coast of Spain between 2012 and 2014, using single or multiple hydrophones. The dataset comprises a total of 90 recordings with a sampling rate of 52734 Hz, and their durations range from 15 seconds to 10 minutes. The total duration of all recordings is approximately three hours. It includes 11 different types of ship sounds and one type of natural noise. Since three categories among them (pilot ships, trawlers, tug boats) lack sufficient samples and durations to support the train-validation-test split, we select a subset of nine types of sounds (dredgers, fish boats, motorboats, mussel boats, ocean liners, passenger ships, ro-ro ships, sailboats, and natural noise) for the main recognition task in this study.

Furthermore, in order to create labels for the auxiliary task, we conduct thorough investigations on all targets in ShipsEar and categorize them into five classes based on their sizes. The mapping relationship between the different types of ships and their corresponding sizes is presented in Table~\ref{tab_map}.

\begin{table*}[htbp]
\normalsize
    \centering
    \caption{The mapping relationship between sizes and types on ShipsEar, with corresponding ranges of length, breadth, draught$^1$.}
	\scalebox{0.8}{\begin{tabular}{lllll}
		\hline
		Size&Type&Length(meters)& Breadth(meters)& Draught(meters)\\
            \hline
            None (no target)&natural noise& -& -& -\\
            Tiny&motorboats, sailboats& 6.2 -- 10.6 & 2.3 -- 3 & 0.45 -- 1\\
            Small&fish boats, mussel boats &11.7 -- 14&$\approx$4&$\approx$1.05\\
            Medium&passenger ships, dredgers$^2$&16 -- 36.5 & 6 -- 10 & 1 -- 2.5\\
            Large&ro-ro ships, ocean liners&164 -- 311.1 & 25.6 -- 49.1 & 6.8 -- 8.6\\
            
		\hline
        \label{tab_map}
	\end{tabular}}
\end{table*}

\begin{tablenotes}
\footnotesize
    \item $1$. The detailed data of length, breadth, draught is mainly acquired from Automatic Identification System (AIS) and \url{http://www.marinetraffic.com.}
    \item $2$. In the ShipsEar article~\cite{santos2016shipsear}, dredgers, fish boats, and mussel boats were grouped into the same category based on size, possibly to maintain a balanced sample quantity for each category. However, through our investigation, it has been found that dredgers (e.g., the length and breadth of the dredger ``Adricristuy'' is 36.48m $\times$ 9m) are more appropriately classified as medium-sized targets, similar to passenger ships. Therefore, this study categorizes dredgers as medium-sized targets instead of small ones.
    
\end{tablenotes}
 

\subsection{Data division}
In this study, each signal recording was divided into segments of 30 seconds, with a 15-second overlap. To prevent any potential information leakage, we ensured that the segments in the training and test sets were not sourced from the same audio track. Furthermore, to ensure a fair comparison, we adopted a previously released train-test split~\cite{xie2024unraveling} for ShipsEar, as displayed in Table~\ref{tabD1}. For this setting, 15\% of the data in the training set is randomly chosen as the validation set. 

\begin{table*}[ht]
\normalsize
    \caption{\label{tabD1} The train-test split for ShipsEar. The ``ID'' in the table refers to the ID of the .wav file in the dataset$^3$.}
    \centering
	\scalebox{0.9}{
	\begin{tabular}{lll}
        \hline
	Category& ID in Training set & ID in Test set\\
	\hline
	Dredger&80,93,94,96& 95 \\
    Fish boat&73,74,76&75 \\
    Motorboat&21,26,33,39,45,51,52,70,77,79&27,50,72 \\
    Mussel boat&46,47,49,66&48 \\
    Natural noise&81,82,84,85,86,88,89,90,91&83,87,92\\
    Ocean liner&16,22,23,25,69&24,71\\
    Passenger ship\quad\quad\quad&06,07,08,10,11,12,14,17,32,34,36,38,40,\quad\quad  
    &9,13,35,42,55,62,65\\
    &41,43,53,54,59,60,61,63,64,67& \\
    RO-RO ship&18,19,58&20,78\\
    Sailboat&37,56,68&57\\
        \hline
	\end{tabular}}
\end{table*}

\begin{tablenotes}
\footnotesize
    \item $3$. This split is also released on https://github.com/xy980523/ShipsEar-An-Unofficial-Train-Test-Split.
\end{tablenotes}

\subsection{Training Parameter Setup}
For training, all models employed the AdamW optimizer~\cite{loshchilov2017decoupled}. The maximum learning rate was set to 5$\times 10^{-4}$, and the weight decay was set to $10^{-5}$ for all experiments. The cosine annealing schedule was adopted for learning rate decay and the warm-up epoch was set to 5. All models were trained for 200 epochs using Nvidia A40 GPUs, with a CUDA version of 11.4. The Python version was 3.8.8 and the Pytorch version is 1.9.0.

\section{Results and Analyses}
In this study, the uniform evaluation metric adopted for the multi-class recognition or estimation tasks is the accuracy rate, which is determined by dividing the number of correctly predicted samples by the total number of samples. Notably, we reported accuracy at the 30-second segment level instead of the record-file level to avoid the occurrence of identical file-level accuracy among multiple groups of experiments due to the limited number of record files in the test set. To mitigate the impact of randomness, each experiment was performed twice using different random seeds (42, 123), and we reported the mean and the unbiased standard deviation (stdev) of the segment-level accuracy produced by the two runs. For instance, we would report 75.60(mean)$\pm$0.84(stdev)\% when the results of two runs conducted with distinct random seeds were 75.00\% and 76.19\%. Furthermore, we uniformly applied the local masking and replicating (LMR) technique~\cite{xu2023underwater}, as proposed by Xu et al., to all our implemented experiments for data augmentation purposes. For M3 and M3-TSE models, the number of expert layers was set to three by default. Specifically, M3 consists of three shared experts, while M3-TSE is composed of one shared expert and two task-specific experts.

\subsection{Preliminary Experiments on Features and Models}

\begin{figure*}
    \centering
    \includegraphics[width=1.0\linewidth]{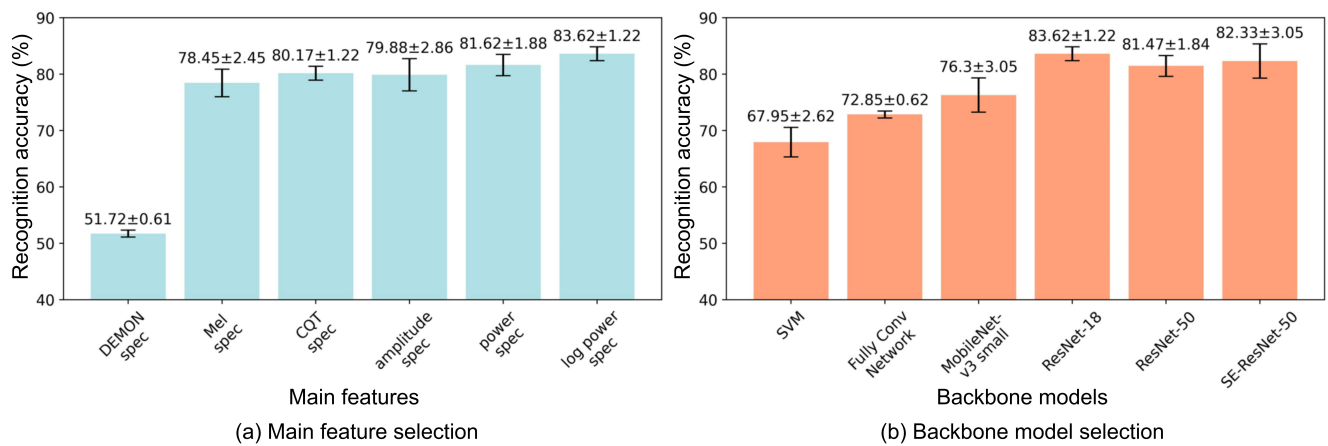}
    \caption{The results of the preliminary experiments conducted to select the optimal main feature and model backbone. The values of error bars represent the unbiased standard deviation between the results of two runs conducted with different random seeds. ``spec'' is the abbreviation for ``spectrogram''.}
    \label{fig_result1}
    \vspace{-2px}
\end{figure*}

\begin{figure*}
    \centering
    \includegraphics[width=1.0\linewidth]{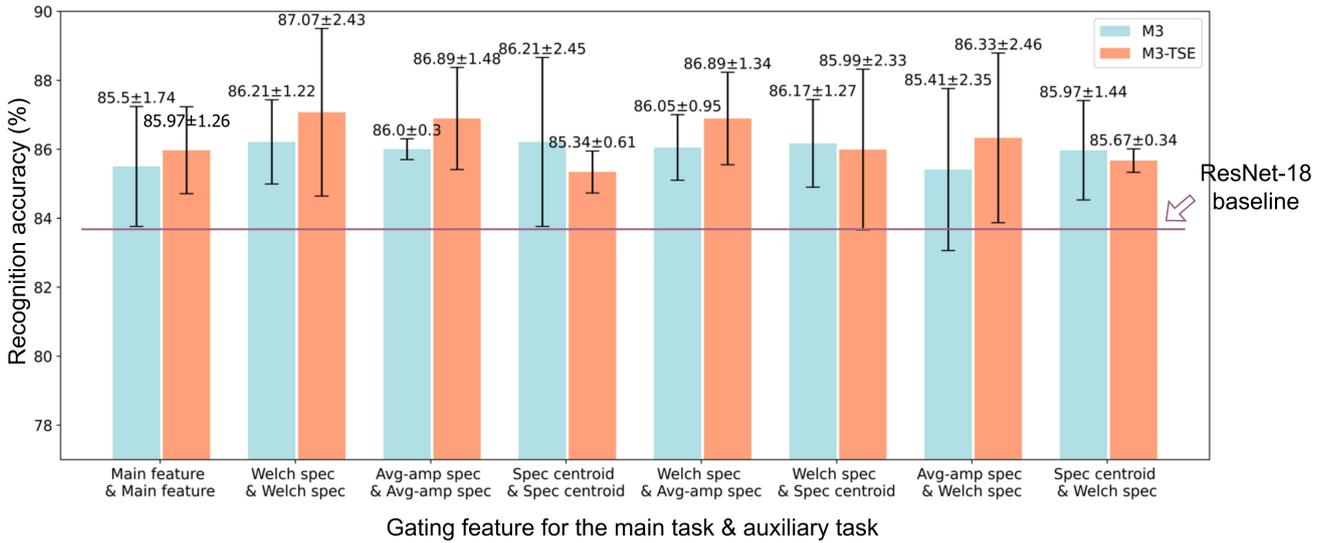}
    \caption{The results of the preliminary experiments conducted to select the optimal gating features for both tasks. The values of error bars represent the unbiased standard deviation between the results of two runs conducted with different random seeds. The scale label on the horizontal axis represents the gating feature for the main task \& auxiliary task. For example, ``Avg-amp spec \& Welch spec'' indicates that the average amplitude spectrum serves as the gating feature for the main task, while the Welch spectrum serves as the gating feature for the auxiliary task.}
    \label{fig_result2}
    \vspace{-2px}
\end{figure*}

First, preliminary experiments were performed to select the optimal main feature, model backbone, and gating feature. These preliminary experiments focused on the 9-class type recognition task. The experiments related to main features and model backbones are presented in Fig.~\ref{fig_result1}, while the experiments regarding the gating features are illustrated in Fig.~\ref{fig_result2}. Fig.~\ref{fig_result1} (a) showcases the performance of models utilizing different main features. To ensure a fair comparison, ResNet-18 was consistently employed as the recognition model. The results indicate that the log power spectrogram outperforms the other candidate features (DEMON spectrum~\cite{lu2020fundamental}, Mel spectrogram~\cite{liu2021underwater,tang2023differential}, CQT spectrogram~\cite{irfan2021deepship,xie2023guiding}, amplitude spectrogram~\cite{xie2024unraveling}, power spectrogram~\cite{yin2020underwater,li2022combined}), achieving an average recognition accuracy of 83.62\%. According to our analysis, the power spectrum, which exhibits stationarity and favorable statistical properties~\cite{nguyen2017modeling}, can capture more comprehensive time-varying patterns and exhibit better noise robustness when processing non-stationary signals~\cite{sha2014underwater}. Additionally, taking the logarithm of the power spectrum can amplify the detailed information of the low-energy part while avoiding dominance by the high-energy part, which may include interference noise such as pulses. Considering the experimental results and analyses, the log power spectrogram can serve as the optimal main feature in this work.

Fig.~\ref{fig_result1} (b) presents the performance of different model backbones. In this set of experiments, the log power spectrogram was uniformly employed as the input feature. Among them, ResNet-18 achieved superior performance compared to other models that have been applied to underwater acoustic recognition, such as support vector machine (SVM)~\cite{yang2016underwater}, fully convolutional network (FCN)~\cite{liu2021underwater}, MobileNet~\cite{akbarian2023recognition}, ResNet-50~\cite{xie2022adaptive,ren2022ualf}, and SE-ResNet~\cite{xue2022novel}. Notably, for the underwater acoustic recognition task, vanilla ResNet-18 outperformed other ResNet-based models, including ResNet-50 and SE-ResNet. This observation suggests that the structure and complexity of ResNet-18 are well-suited for the task and can serve as a reliable model backbone. In this study, M3 adopts the architecture of ResNet-18 for its various modules, where the expert layer conforms to the front-end layers of ResNet-18, and the tower network exactly follows the four residual layers of ResNet-18.

Furthermore, to evaluate the advantages of gating features (Welch spectrum, average amplitude spectrum, spectral centroid) over traditional ones (main feature), this study compared the performance of M3 (M3-TSE) using different gating features, as depicted in Fig.~\ref{fig_result2}. The main task and the auxiliary task can share the same gating feature or adopt distinct gating features. Firstly, our experimental findings indicate that, following the strategies employed in traditional MoE approaches, utilizing the main feature as the gating feature mostly results in inferior performance compared to employing 1-D frequency domain features. This suggests that non-redundant, low-dimensional features are more appropriate for providing task-relevant knowledge to the gating layer. Although high-dimensional main features may contain more information, their complexity and redundancy contribute to additional computational costs and may lead the gating layer to overfit. Furthermore, the results also demonstrate that employing Welch spectra as the gating feature for both tasks yields the optimal outcomes for M3 or M3-TSE, corroborating the analysis of feature distributions depicted in Fig.~\ref{fig_gate_input}. The statistical properties of Welch spectra provide them with enhanced stability when dealing with non-stationary underwater acoustic signals, as well as more discernible and resilient patterns. Therefore, Welch spectra are the most suitable choice to serve as the gating feature.

\subsection{Main Results}

\begin{table*}[htbp]
\normalsize
    \centering
    \caption{The main results on the Shipsear dataset, including the comparative experiments between our methods and current advanced methods. ``Type Acc'' indicates the accuracy of the 9-class type recognition task, while ``Size Acc'' indicates the accuracy of the 5-class target size estimation task. The $(\cdot)$ in the ``Parameter Number'' column represents the number of parameters that can be conditionally pruned.}
	\scalebox{0.84}{\begin{tabular}{llcccc}
		\hline
		 &Method& Parameter Number&Type Acc(\%)&Size Acc(\%)\\
            \hline
        Current advanced methods\quad &Official benckmark~\cite{santos2016shipsear}&-& -&75.40\\
        &UALF~\cite{ren2022ualf}&-&-&80.73\\
        &AGNet~\cite{xie2022adaptive}&3.0025$\times10^{7}$&-&85.48\\
        &Smoothness Regularization \& LMR~\cite{xu2023underwater}&1.1175$\times10^{7}$&83.45 &-\\
        &I-Contrastive Learning~\cite{xie2023guiding}&2.2870$\times10^{7}$&85.34&-\\
        &CMoE$^\ast$~\cite{xie2024unraveling}&1.1190$\times10^{7}$&86.21&-\\
        \hline
        Our methods&ResNet-18 (single-task baseline)&1.1175$\times10^{7}$&83.62$\pm$1.22&87.21$\pm$0.00\\
        

        &Fundamental multi-task model ($\alpha=1.0$)&1.1178(+1.1169)$\times10^{7}$&85.34$\pm$0.74 &87.07$\pm$0.61\\
        &M3($\alpha=1.0$)&1.1347(+1.1338)$\times10^{7}$&86.21$\pm$1.22&87.93$\pm$1.22\\
        &M3-TSE($\alpha=2.0$)&1.1347(+1.1338)$\times10^{7}$&\textbf{87.07$\pm$2.43} &\textbf{90.52$\pm$1.23}\\
        
        \hline
        
        \label{tab3}
        \vspace{-2mm}
	\end{tabular}}
 
 \begin{tablenotes}
 \footnotesize
    \item $\ast$ According to our investigation of previous work, CMoE~\cite{xie2024unraveling} with an accuracy of 86.21\% serves as the state-of-the-art on the 9-class target type recognition task.
    
 \end{tablenotes}

\end{table*}

In this subsection, we present the main experimental results in Table~\ref{tab3}, which include comparisons with our M3 models and other advanced methods~\cite{xie2023advancing, xie2022adaptive, xie2023guiding, ren2022ualf}. Firstly, our results clearly demonstrate the superior performance of multi-task models on both tasks. On the 9-class type recognition task, the fundamental multi-task model (see Fig.~\ref{fig2}(a)) shows a 1.72\% performance improvement compared to its corresponding single-task baseline. It can also achieve competitive recognition accuracy compared to the most advanced single-task models~\cite{xie2023guiding,xie2024unraveling}. This supports the necessity of transitioning from single-task to multi-task paradigm. Whereas, the fundamental multi-task model exhibits a slight performance decrease of 0.14\% in the 5-class size estimation task compared to its single-task counterpart. This is due to the shared layer lacking an independent parameter space, resulting in mutual interference between tasks. Nevertheless, the incorporation of multiple experts and gating mechanisms within M3 and M3-TSE can effectively address this issue. Notably, M3-TSE outperforms the fundamental multi-task model with performance gains of 1.73\% and 3.45\% on respective tasks, reaching state-of-the-art performance in the 9-class type recognition task. M3-TSE operates within independent parameter spaces to handle intricate underwater acoustic signals and leverages the task-specific expert mechanism to further alleviate inter-task interference.

Moreover, Table~\ref{tab3} also presents the number of trainable parameters for each model. M3 and M3-TSE are relatively complex models, typically consisting of 2.2685$\times10^{7}$ parameters. However, during practical model deployment for inference or testing, it is optional to prune the gating layer and tower network used for the auxiliary task, while retaining the remaining modules. This approach can effectively reduce the model's parameter count to 1.1347$\times10^{7}$, albeit sacrificing the capacity to estimate target sizes. In general, the M3 structure offers a certain level of flexibility to adapt to various application requirements. It allows for simultaneous prediction results of multiple tasks, and can also conditionally perform efficient inference with fewer parameters and lower computational consumption.

\begin{table*}[htbp]
\normalsize
    \centering
    \caption{The results of the ablation experiments. The results of models with TSE mechanism are obtained with $\alpha=2.0$, while the results of the other models are achieved with $\alpha=1.0$.}
	\scalebox{0.8}{\begin{tabular}{cccccc}
		\hline
	\quad Multi-task\quad &\quad Multi-expert(gate)\quad & \quad 1-D gating feature\quad &\quad TSE mechanism\quad &\quad Type Acc(\%)\quad &\quad Size Acc(\%)\quad\\
            \hline
        $\times$ & $\times$ &$\times$ & $\times$ & 
         83.62$\pm$1.22  &  87.21$\pm$0.00 \\
        $\checkmark$ & $\times$ &$\times$ & $\times$ & 
         85.34$\pm$0.74  &  87.07$\pm$0.61 \\
        $\checkmark$ & $\checkmark$ &$\times$ & $\times$ & 
         85.50$\pm$1.74  &  87.21$\pm$1.26 \\
         $\checkmark$ & $\checkmark$ &$\times$ & $\checkmark$ & 
         85.97$\pm$1.26  &  87.93$\pm$0.00 \\

         $\checkmark$ & $\checkmark$ &$\checkmark$ & $\times$ & 
         86.21$\pm$1.22  &  87.93$\pm$1.22 \\
         $\checkmark$ & $\checkmark$ &$\checkmark$ & $\checkmark$ & 
         87.07$\pm$2.43  &  90.52$\pm$1.23 \\
         
        \hline
        
        \label{tab_abl}
        \vspace{-2mm}
	\end{tabular}}
\end{table*}

Furthermore, to examine the contributions of each module to the promising results, ablation experiments were implemented as depicted in Table~\ref{tab_abl}. The results show that all the modules proposed in this study can improve the recognition results, particularly the multi-expert and multi-gating techniques, which offer nearly risk-free and stable performance improvements. Regarding the TSE technique, although the results in Fig.~\ref{fig_result2} suggest that TSE can have some negative effects in certain cases, this is because TSE may lead to a decrease in the model's ability to capture shared patterns across tasks. However, the results in Table~\ref{tab_abl} demonstrate that when experimental configurations (such as suitable gating features and alpha values) are set appropriately, TSE still exhibits a notable positive effect and can achieve higher performance upper bound. Furthermore, the outcomes presented in Table~\ref{tab_abl} serve as evidence to substantiate the complementarity of the modules proposed in this study, as their cohesive application can bring superior results.


\subsection{Weight Coefficient Between Tasks}


\begin{table}[htbp]
    \centering
    \caption{The comparative experiment results of M3 and M3-TSE using different weight coefficient $\alpha$. Dynamic $\alpha$ indicates the usage of the homoscedastic uncertainty algorithm to automatically learn the value of $\alpha$.}
    \scalebox{0.69}{\begin{tabular}{lcccc}
    \hline
          &\multicolumn{2}{c}{9-class type accuracy}&  \multicolumn{2}{c}{5-class size accuracy}
         \\\cmidrule(lr){2-3}\cmidrule(lr){4-5}
         &\quad\quad M3\quad\quad& \quad\quad M3-TSE \quad\quad & \quad\quad M3 \quad\quad & \quad\quad M3-TSE \quad\quad\\
         \midrule
         $\alpha$=0.1 \quad\quad& 84.05$\pm$0.61&83.62$\pm$2.43&85.34$\pm$1.84&87.21$\pm$0.00\\

        $\alpha$=0.2 \quad\quad& 85.35$\pm$2.45&85.35$\pm$1.23&85.78$\pm$1.84&88.93$\pm$1.22\\

        $\alpha$=0.5 \quad\quad& 86.21$\pm$0.61&84.48$\pm$1.22&86.21$\pm$2.43&87.21$\pm$1.23\\

        $\alpha$=1.0 \quad\quad& \textbf{86.21$\pm$1.22}&87.07$\pm$1.84&\textbf{87.93$\pm$1.22}&88.93$\pm$0.00\\

        $\alpha$=2.0 \quad\quad& 85.78$\pm$0.62&\textbf{87.07$\pm$2.43}&87.93$\pm$0.61&\textbf{90.52$\pm$1.23}\\

        $\alpha$=5.0 \quad\quad& 84.34$\pm$0.95&84.48$\pm$1.22&87.07$\pm$0.61&87.21$\pm$0.00\\

        \hline
        Learnable $\alpha$ \quad\quad& 85.48$\pm$1.22&86.64$\pm$0.30&87.21$\pm$0.62&89.66$\pm$1.23\\
        
         \hline
         
    \end{tabular}}
    
    \label{tab:ALPHA}
\end{table}

The weight coefficient $\alpha$ between tasks, as defined in Equation (3), also plays a crucial role in the performance of M3 and M3-TSE. By adjusting the value of $\alpha$, the weight assigned to each task in the loss function can be modified, thereby influencing the model's preference. In addition to manually setting several candidate values ($\alpha\in$\{0.1, 0.2, 0.5, 1, 2, 5\}), we also employed the homoscedastic uncertainty algorithm~\cite{kendall2018multi,neilsen2021learning} to learn the optimal $\alpha$ value in an adaptive manner. In this algorithm, $\alpha$ was converted into a learnable parameter with an initial value of 1.

The results of M3 and M3-TSE across different values of $\alpha$ are presented in Table~\ref{tab:ALPHA}. Notably, setting $\alpha$ to 1 or 2 proves to be a valid choice for both models. A moderate value of $\alpha$ ensures a balanced weighting between tasks, thereby assisting in mitigating inter-task conflicts. Conversely, excessively large or small weights (e.g., $\alpha$=0.1 or 5) can potentially disrupt the balance between tasks, leading to one task dominating the model's training process and another task being overly regularized. This imbalance may compromise the model's ability to capture shared characteristics or patterns across tasks, resulting in a decline in its capacity to handle both tasks effectively. Besides, although models utilizing the homoscedastic uncertainty algorithm do not achieve optimal results, the automatic learning of $\alpha$ still provides valuable insights. The learnable $\alpha$ values ultimately converge to approximately 0.911 for M3 and 0.933 for M3-TSE, further supporting the claim that the model favors a moderate value for $\alpha$ and a balanced weight between tasks.

\section{Conclusion}
This study begins by identifying the limitations of current acoustic recognition methods and proposes the utilization of a multi-task framework to leverage the inherent properties of targets. To fully realize the potential of the multi-task paradigm, this study adopts multi-expert and multi-gate mechanisms, which process complex underwater acoustic signals using specialized and separate parameter spaces. Additionally, 1-D frequency-domain features are incorporated as gating features, and certain expert layers are transformed into task-specific ones. Through comprehensive comparative and ablation experiments, the superior efficacy of M3 and ME-TSE is validated on the ShipsEar dataset.

Despite achieving state-of-the-art results on the ShipsEar dataset, this study has certain limitations. For example, due to the limited scale of the ShipsEar dataset, we did not investigate the MoE structure with a larger number of experts, such as 8, 16, or even 128. When dealing with large-scale datasets that encompass diverse feature spaces, the current 3-expert M3 model may not fully exploit the potential of the MoE structure. Furthermore, there is room for improvement in the strategy for constructing the auxiliary task. Although the strong correlation between target size estimation and the recognition task helps mitigate interference between tasks, the auxiliary task provides limited supplementary information, thereby offering limited benefits to recognition systems. In future research, we intend to explore more innovative designs for auxiliary tasks to push the performance boundaries of multi-task systems to their upper limits.

\section*{Acknowledgements}
This research was partially supported by the Chinese Academy of Sciences Strategic Leading Science and Technology Project (No. XDA0310103); the IOA Frontier Exploration Project (No. ZYTS202001); and the Youth Innovation Promotion Association CAS.

\section*{Author Declarations}
\textbf{Conflict of Interest} The authors declare that they have no known competing financial interests or personal relationships that could have appeared to influence the work reported in this paper.

\section*{Data Availability}
The ShipsEar dataset that supports the findings of this study is openly available at: \url{https://underwaternoise.atlanttic.uvigo.es/}; Our train-test-split is available at: \url{https://github.com/xy980523/ShipsEar-An-Unofficial-Train-Test-Split}.

\section*{Declaration of Generative AI and AI-Assisted Technologies in the Writing Process}
During the preparation of this work, we used $ChatGPT$ in order to check the English grammar and word spelling. After using this service, we reviewed and edited the content as needed and took full responsibility for the content of the publication.


\newpage

\printcredits

\bibliographystyle{cas-model2-names}
\bibliography{cas-refs-M3}





\end{document}